\documentclass[%
 aip,
 jmp,%
 amsmath,amssymb,
preprint,%
]{revtex4-1}

\usepackage{graphicx}
\usepackage{dcolumn}
\usepackage{bm}

\begin{document}

\preprint{AIP/123-QED}

\title[]{Maze Solving Automatons for Self-Healing of Open Interconnects: Modular Add-on for Circuit Boards}
\thanks{Corresponding Author: Sanjiv Sambandan, Assistant Professor, Department of Instrumentation and Applied Physics, Indian Institute of Science, Malleswaram, Bangalore 560012, India. Ph: +91-80-22933196. email: sanjiv@iap.iisc.ernet.in, ssanjiv@isu.iisc.ernet.in.}

\author{Aswathi Nair}
\affiliation{Department of Instrumentation and Applied Physics, Indian Institute of Science, Bangalore, India}

\author{Karthik Raghunandan}%
\affiliation{Department of Instrumentation and Applied Physics, Indian Institute of Science, Bangalore, India}

\author{Vaddi Yaswant}
\affiliation{Department of Instrumentation and Applied Physics, Indian Institute of Science, Bangalore, India}

\author{Sreelal Shridharan}
\affiliation{Vikram Sarabhai Space Center, Indian Space Research Organization, Trivandrum, India}

\author{Sanjiv Sambandan}

\affiliation{Department of Instrumentation and Applied Physics, Indian Institute of Science, Bangalore, India}

\date{\today}

\begin{abstract}
We present the circuit board integration of a self-healing mechanism to repair open faults. The electric field driven mechanism physically restores fractured interconnects in electronic circuits and has the ability to solve mazes. The repair is performed by conductive particles dispersed in an insulating fluid. We demonstrate the integration of the healing module onto printed circuit boards and the ability of maze solving. We model and perform experiments on the influence of the geometry of the conductive particles as well as the terminal impedances of the route on the healing efficiency. The typical heal rate is 10 $\mu$m/s with healed route having resistance of 100 $\Omega$ to 20 k$\Omega$ depending on the materials and concentrations used.
\end{abstract}

\maketitle

Open interconnect faults between inter layer vias, solder joints and routes are causes of poor reliability of printed circuit boards (pcbs). Several factors such as electro-migration, electro-static discharge, mechanical and thermal stress combined with poor fabrication are the cause of such faults \cite{1}-\cite{6}. The advent of organic electronics permits integrated circuits on large area mechanically flexible substrates where long interconnects are prone to open circuit faults \cite{7}. One means to improve the reliability of electronic circuits is by replicating important functional blocks. However, making allowances in track width for all interconnects and vias in high density pcbs becomes expensive in terms of cost, layout area and weight. Therefore integrating a real time repair mechanism that physical restores open routes has advantages. 

There have been several reports on the online repair of open interconnects \cite{8}-\cite{15}. Modulation of transmission line impedances was reported for high frequency application where active circuits sensed changes in impedance and offered correction to achieve best performance \cite{9}. With regards to healing a fatal open fault, encapsulating conductive fluids in dielectric shells have been considered. These shells were embedded in interconnects and broke open to spill out the fluid during the fracture of the route  \cite{10}-\cite{15}. This technique was demonstrated for a range of encapsulants from carbon nanotubes \cite{11}, conductive salts \cite{12}, graphene \cite{13}, carbon black \cite{14} and Ga-In metal \cite{15}.

Here we develop an intelligent, modular and flexible patch that can be added onto pcbs to enable self healing. The intelligence of the repair lies in its ability to solve mazes and heal open interconnects that are not in the line of sight. In this regard a suitable technique involves the field induced aggregation of conductive particles dispersed in an insulating fluid with the dispersion isolated over routes \cite{16},\cite{17}, \cite{18}. Since every route is connected to the power supply and ground via some impedance, an electric field develops in the gap upon the fracture of the route and polarizes the conductive particles. The polarized particles experience dipole interaction resulting in them chaining up to bridge the gap thereby performing repair [Supplementary Video1]. Figure 1 shows the bridging of a 200 $\mu$m gap with a 30 V drop by 10 $\mu$m spherical Cu particles dispersed in Silicon oil. The effectiveness of self heal is determined by the time taken to repair the fault and the current carrying capacity of the repaired interconnect. Some of the factors this depends on are the concentration of the dispersion, permittivity and viscosity of the fluid, conductivity and geometry of the conductive particle, electric field strength and gradient, and the impedance of the peripheral circuits connected to the routes.

\begin{figure}
\includegraphics[width=2.5 in]{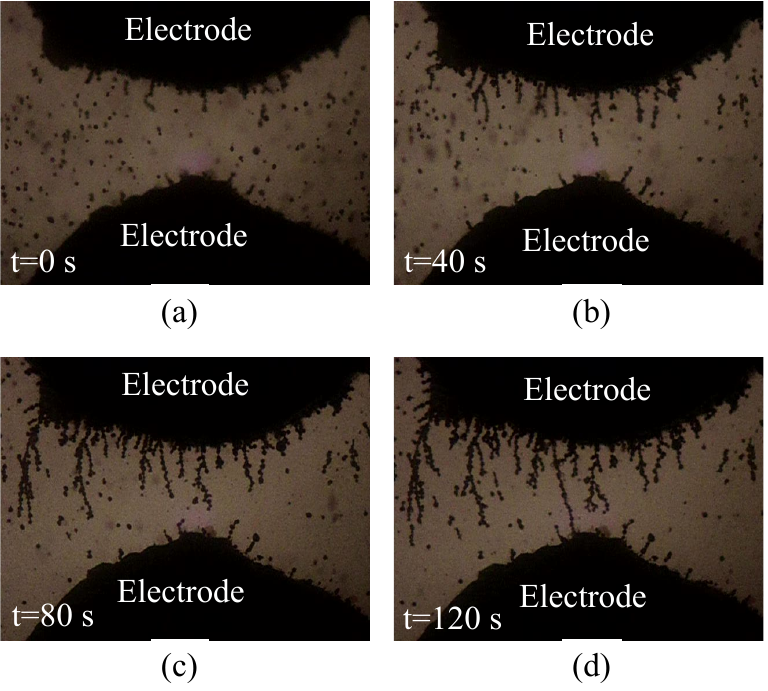}
\caption{(a)-(d) Self-healing of an open interconnect by a conductive bridge formed due to dipole interactions.}
\end{figure}

The role of geometry of conductive particles in the dispersion (in silicon oil) was studied with particles shaped as rods - metallic carbon nanotubes (CNTs) of 10 $\mu$m length, spheres - Cu of 10 $\mu$m diameter, and composite mixture of rods and spheres (CNTs and Cu). Upon polarization, the interaction between two rods requires time for both rotation and translation, while spheres experience only translation. On the other hand, once aligned, the rods experience smaller drag as compared to spheres. Figure 2a shows the current through the interconnect at different field strengths using 0.8 mg/ml CNT dispersion, 100 mg/ml Cu dispersion, and a composite dispersion of 0.5mg/ml CNT-100mg/ml Cu. The current was measured using a Keithley 2636A SMU while keeping the voltage across the 200 $\mu$m gap constant. At time zero, the inter-connect is open. With time, several bridges form across the gap allowing the current to gradually climb till the current saturates at a maxima (compliance current of SMU). Figure 2b shows the variation in current for various dispersion concentrations of the dispersion with the voltage kept constant at 30 V for CNTs, 27 V for Cu, and 25 V for the composite CNT-Cu. CNTs generally achieved and maintained larger currents while spherical Cu particles could not sustain current at high fields (Figure 2b).  Figure 2c shows typical bridges formed between the electrodes.

\begin{figure*}
\includegraphics[width=5 in]{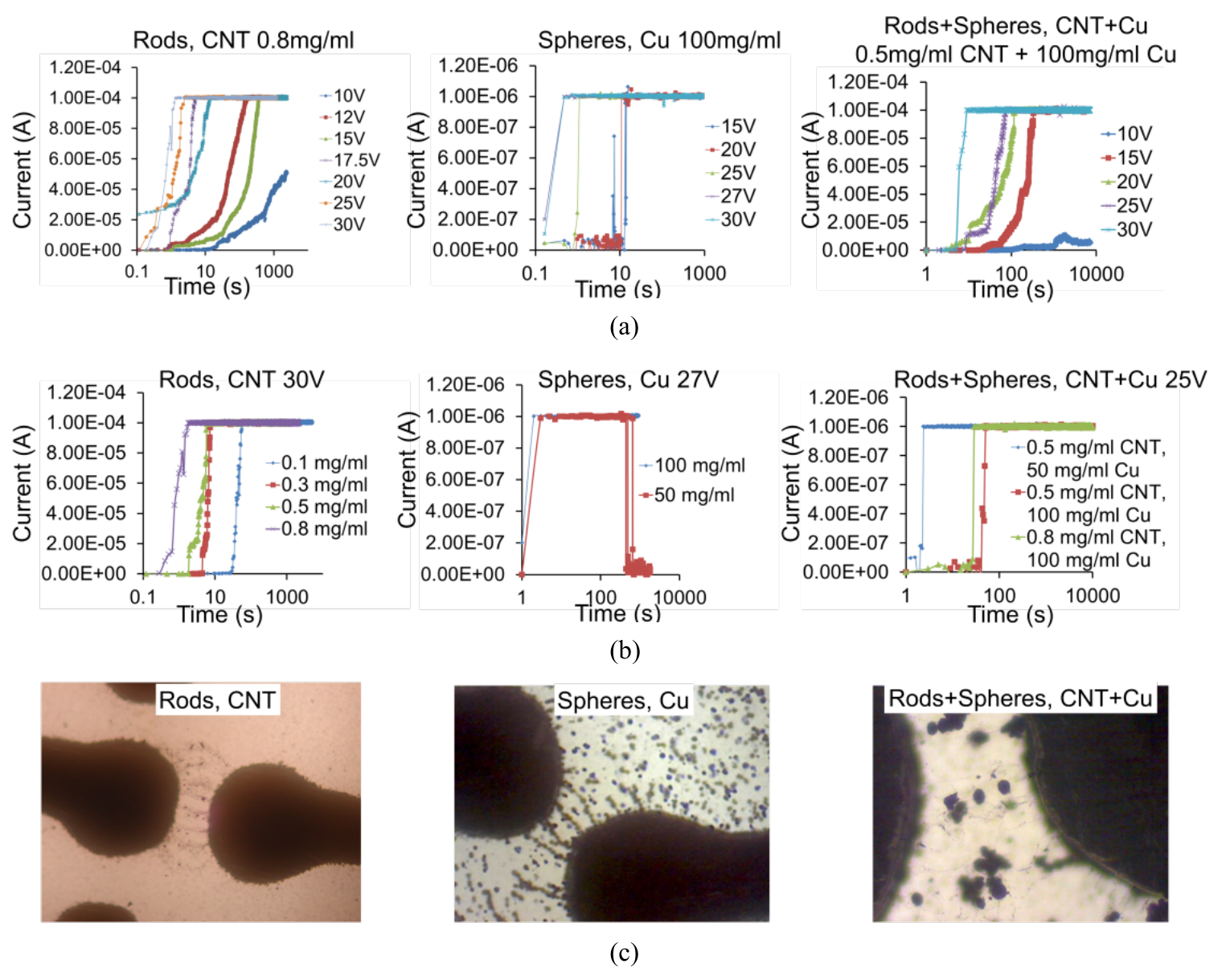}
\caption{Dependence on the shape of conductive particles. Bridging time as a function of (a) electric field, (b) dispersion concentration. (c) Typical bridge profiles after heal.}
\end{figure*}

Ignoring constant coefficients, the rotation of the rod is governed by $\eta R_{R}L^{2}\dot{\theta}=p\xi \mbox{sin}(\theta)$ with $\eta$ being the viscosity of the fluid, $\theta$ the angle between the axis of the rod and the field lines, $R_{R}$ and $L$ the radius and length of the rod, respectively, $p$  the dipole moment and $\xi$ the magnitude of the electric field. The time required for the rotation and alignment of the rod to the field lines varies as $~ (\eta/\epsilon_{f})\xi^{-2}$ where $\epsilon_{f}$ is the permittivity of the isulating fluid. Ignoring constant coefficients, the translation of a sphere in the presence of viscous drag is governed by $\eta R_{s} \dot{x}=p^{2}\epsilon_{f}^{-1}x^{-4}$, where $R_{S}$ is the radius of the sphere and $x$ the distance between two polarized spheres. The time for translation varies as $~(\eta/\epsilon_{f})(d_{0}/R_{S})^{5}\xi^{-2}$ where $d_{0}$ is the average initial spacing between the spheres in a homogenous dispersion and depends on the concentration. In essence, the time taken to conduct repair varies as $\lambda \xi^{-2}$  where the coefficient $\lambda$  is proportional to $\eta/\epsilon_{f}$. This dependence is observed in Figure 3 that shows the time taken to bridge the gap as a function of the electric field for various dispersions.

\begin{figure}
\includegraphics[width=3 in]{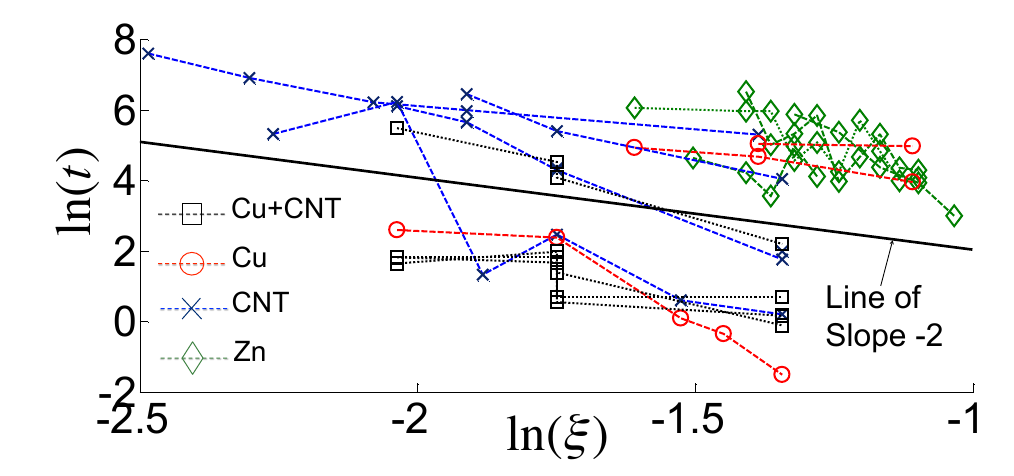}
\caption{The repair time varies as the inverse square of the electric field.}
\end{figure}

There exists a lower and upper bound on the voltage across the gap for proper operation of the self-healing mechanism. If the field is low and below a threshold field, $\xi_{th}$, the time taken to bridge the gap is large. For a large field due to a large applied voltage, the bridge forms rapidly, but is destroyed once formed by the current surge [Supplementary Video2]. Moreover, the voltage across the gap during repair is a function of the terminal impedances and varies with time during repair. Consider a healthy route having input and output impedances $Z_{in}$ and $Z_{out}$, respectively (Figure 4a). Upon fracture of this route, a field $\xi_{0}=V_{0}/D$ develops across the gap of length $D$ and repair is initiated. The time taken for the formation of the first bridge is $\lambda \xi_{0}^{-2}$. If the impedance of the first bridge is $Z_{b}$, the field across the gap reduces to $\xi_{0}(\alpha+1)^{-1}$ where $\alpha=(Z_{in}+Z_{out})/Z_{b}$. The time taken for the formation of the second bridge is larger due to the lower field and is given by $\lambda\xi_{0}^{-2}(\alpha+1)^{2}$. The second bridge results in the healed route having an impedance of $Z_{b}/2$ therefore reducing the field further. This successive formation of bridges and reduction in field continues until the field reaches $\xi_{th}$. The number of bridges that form is $m=\lfloor{(\xi_{0}/\xi_{th}-1)\alpha^{-1}}\rfloor$ and the total time required for this is $T_{b}=\lambda \xi_{0}^{-2}\sum_{j=0}^{j=m}(j\alpha+1)^{2}$. The effective impedance of, and current through the repaired route would be $Z_{b}/m$ and $I_{b}=(V_{0}/Z_{b})(\alpha+m^{-1})^{-1}$. Experiments to measure the healing time using different values of equal terminal impedances ($Z_{in}=Z_{out}$) were conducted using 100 mg/ml Cu particles and 50 V applied voltage. From the steady state current through the healed line the bridge resistance was calculated for each case. For zero terminal impedance, the current limited by the heal alone is $mV_{0}/Z_{b}$. The ratio of the repair time and this current is shown in Figure 4b as a function of the external impedance used.

\begin{figure}
\includegraphics[width=2.5 in]{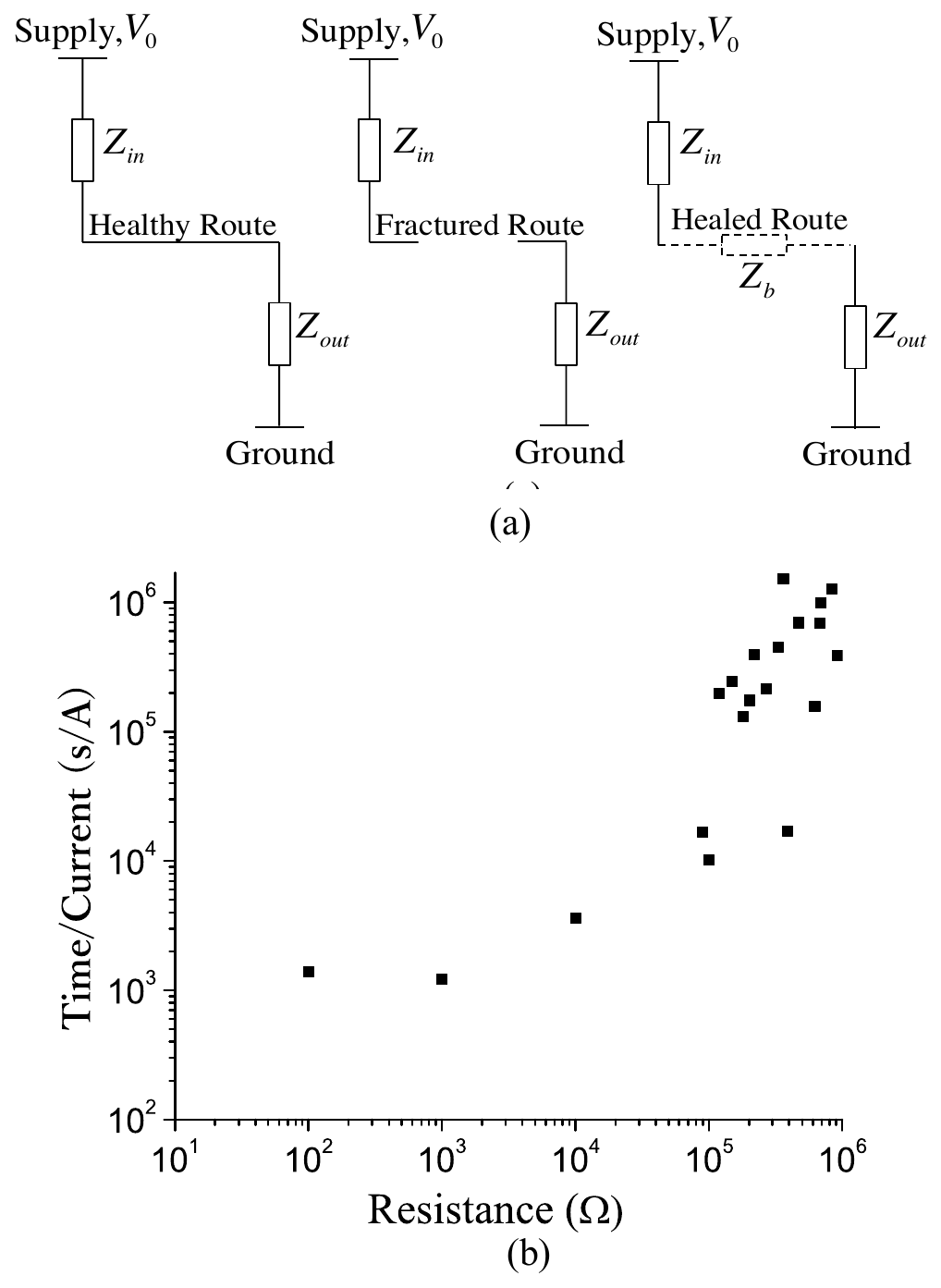}
\caption{Dependence of repair time on the input and output impedances. (a) Repair in the presence of terminal impedances. (b) Repair time to current ratio as a function of terminal impedance ($Z_{in}=Z_{out}$).}
\end{figure}

For integration onto pcbs, the dispersion has to be isolated over the routes. This is achieved as a modular add on that does not change the existing pcb manufacturing technology. From the circuit layout, a template is created with raised features of the routes. A poly-dimethyl-silicoxane (PDMS) mat is molded to this template and overlaid on the pcb thereby creating conformable vesicles overlapping the routes. The dispersion is then injected into these vesicles (Figure 5a). Figure 5b shows the testing of the pcb with a purposefully created open fault. Upon power up, the initially absent output of the amplifier begins to show after healing is complete.

The dispersion concentration determines the parasitic impedances between adjacent routes. As the concentration of the homegenous dispersion is increased, the disperion shifts from being mostly insulating (high impedance) to mostly conducting (low impedance) with a critical threshold occuring with the presence of a percolative conductive path. For dispersions having low impedance, the signal in a route is strongly coupled to an adjacent route via the dispersion thereby increasing parasitic impedances. This results in poor signal to noise ratio and lower bandwidth. Therfore there exists a trade off between improved healing efficiency and increased parasitics due to an increase in dispersion concentration.

\begin{figure}
\includegraphics[width=2.5 in]{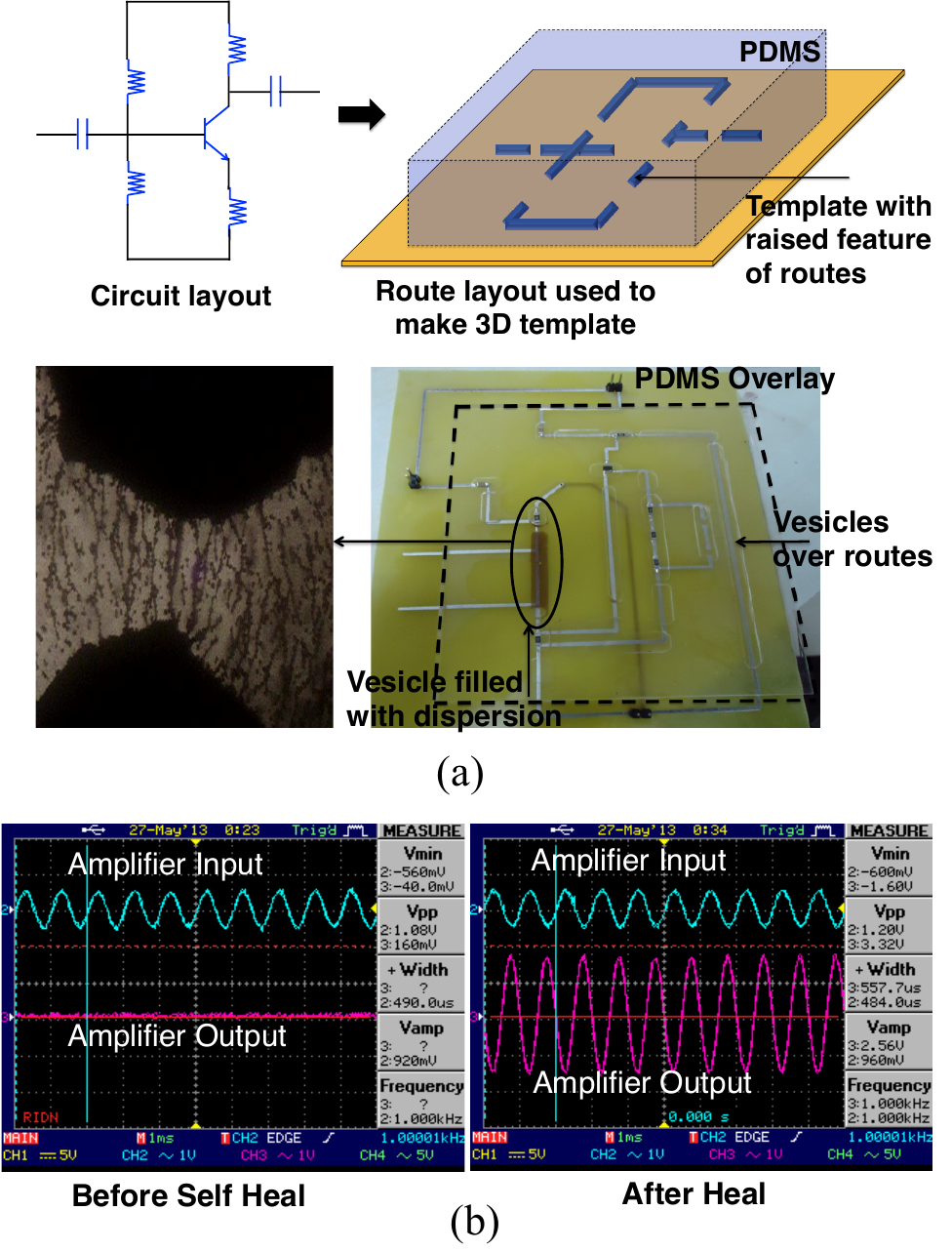}
\caption{(a) Integration of self healing mechanism with an electronic pcb. (b) Demonstration of healing on a common emitter amplifier circuit. A fault is purposefully created in the bias circuit. The amplifier output is initially absent, but appears after repair.}
\end{figure}

The mechanism of self-healing has an inherent ability to solve mazes [Supplementary Video 3]. 2D mazes with high sidewalls were constructed in PDMS with entry and exit points defined with electrodes (Figure 6). Silicon oil was filled to a shallow level contained within the maze. Then a drop of the dispersion added at the entry was allowed to diffuse for some time. When a voltage was applied across the electrodes the dispersion bridged the entry to the exit (solving the maze) and did not diffuse beyond the exit and first chose the most direct path before attempting several other routes. As the size of the maze layout was of the order of cm, voltages of 1000 V (Figure 6a) and 5500 V (Figure 6b) were applied to demonstrate the phenomena. The ability to solve mazes has a simple argument related to the strength of the electric field lines along the solution to the maze (Figure 6c). Figure 6d shows the measured current between the entry and exit for the mazes. Perhaps a more general thermodynamic argument lies in the maximum entropy production principle (MEPP) \cite{19}. Given a potential drop between electrodes, MEPP demands the entropy production be maximized by the creation of a conductive path so as to maximize current and reduce the potential drop as rapidly as possible. This ability of the dispersion to bridge the two electrodes around obstacles aids the repair of routes around debris as shown in Figure 6e. 

\begin{figure}
\includegraphics[width=3 in]{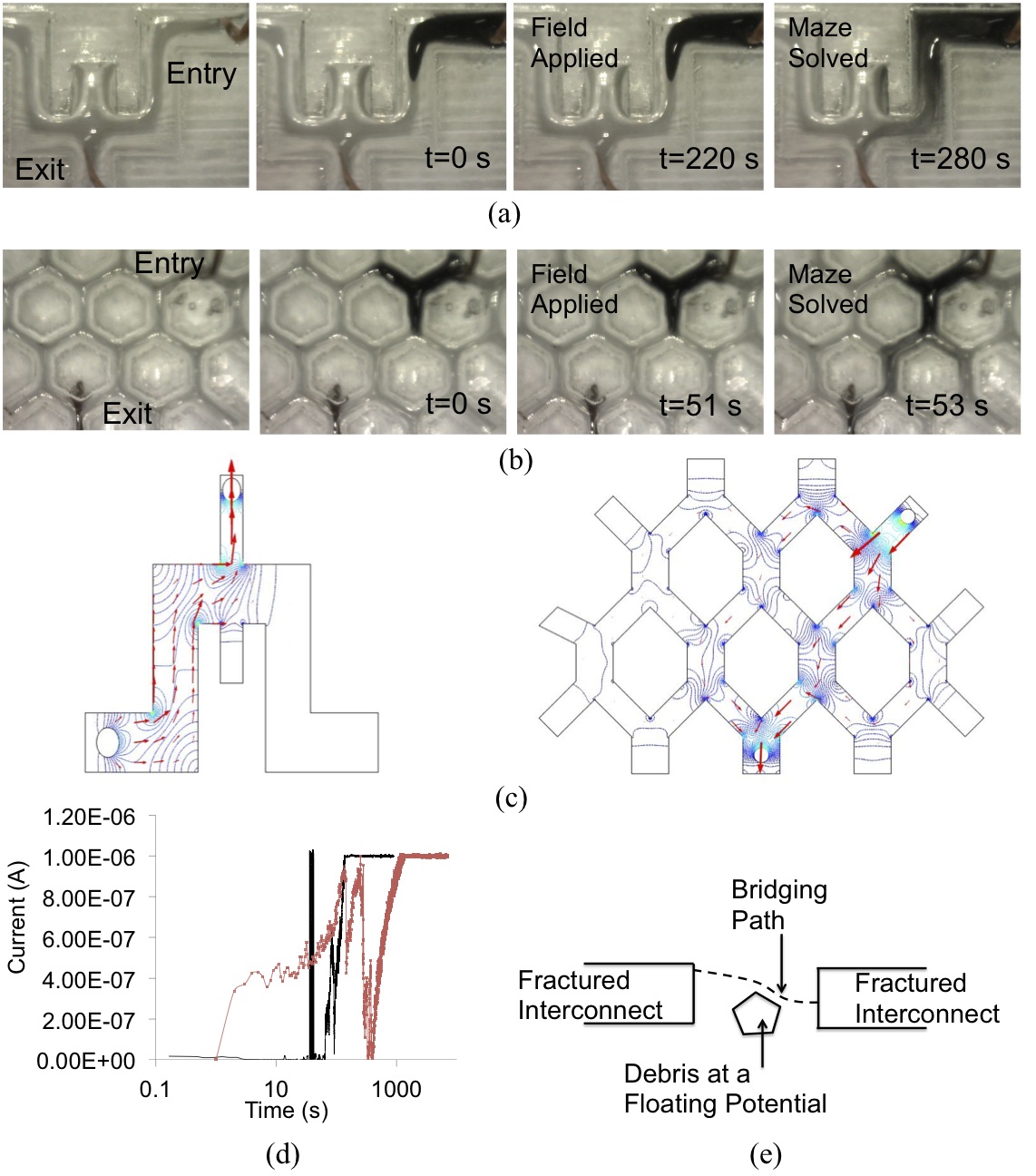}
\caption{Solving a maze. (a) Maze with a unique path to exit. The dispersion is allowed to diffuse till t=220s. When 1000 V is applied the dispersion rapidly finds the exit. (b) Maze with multiple paths to exit. The dispersion is allowed to diffuse till t=51s. When 5500 V is applied the dispersion rapidly finds the exit. [Supplementary Video 3]. (c) Electric field lines in both cases. (d) Current variation with time during maze solving. (e) Find paths around debris.}
\end{figure}

One could call this mechanism a thermodynamic automaton as it remains inert until the occurrence of the fault. The occurrence of the fault triggers the motive force to initiate repair. The movement and activity to repair the fault is the response to this force. The force naturally dies out once repair is satisfactorily complete. The modular add on to repair pcbs has applications for weight assignment in neural networks and space technology where faulty boards on satellites can be restored without the need for expensive retrieve and repair operations.

The authors acknowledge the Indian Space Research Organization for funding received via the Space Technology Cell (Grant No. STC/0298).

\end{document}